\documentclass[]{siamart190516}
\usepackage{amsmath,amsfonts,amssymb,bm,hyperref,xcolor,amsopn}
\usepackage{graphicx,subfigure,epstopdf,multirow,diagbox,caption,algorithmic,enumerate}
\usepackage{xcolor}
\usepackage{soul}

\newcommand\bx{\boldsymbol{x}}

\newcommand\bD{\boldsymbol{D}}

\newcommand\ba{\boldsymbol{a}}
\newcommand\bP{\boldsymbol{P}}
\newcommand\bE{\boldsymbol{E}}

\newcommand{\TheTitle}{Phase Field Formulation in Ferroelectric Thin Film}
\newcommand{\TheAuthors}{Qiang Du, Ruotai~Li, and Lei Zhang}

\headers{\TheTitle}{\TheAuthors}
\title{{\TheTitle}\thanks{This work was funded by the National Natural Science Foundation of China No. 11622102, 11861130351.}}

\author{Qiang Du\thanks{Department of Applied Physics and Applied Mathematics and Data Science Institute, Columbia University, New York, NY 10027, USA (\email{qd2125@columbia.edu}).}
\and Ruotai~Li\thanks{Beijing International Center for Mathematical Research, Peking University, Beijing 100871, China (\email{liruotai@pku.edu.cn}).}
\and  Lei Zhang\thanks{Beijing International Center for Mathematical Research, Center for Quantitative Biology, Peking University, Beijing 100871, China (\email{zhangl@math.pku.edu.cn}).}
}
\ifpdf
\hypersetup{
  pdftitle={\TheTitle},
  pdfauthor={\TheAuthors}
}
\fi
\renewcommand{\TheTitle}{Variational Phase Field Formulations of Polarization and Phase Transition in Ferroelectric Thin Films}

\begin{document}

\maketitle

\begin{abstract}
Electric field plays an important role in ferroelectric phase transition. There have been numerous phase field formulations attempting to account for electrostatic interactions subject to different boundary conditions. In this paper, we develop new variational forms of the phase field electrostatic energy and the relaxation dynamics of the
polarization vector that involves a hybrid representation in both real and Fourier variables. The new formulations
avoid ambiguities appeared in earlier studies and lead to much more effective ways to perform variational studies and
numerical simulations. Computations of polarization switching in a single domain by applying the new formulations are provided as illustrative examples.
\end{abstract}



\section{Introduction}
Ferroelectrics, first discovered in 20th century, are materials possessing a spontaneous polarization that can be switched between energetically equivalent states in a single crystal by an electric field \cite{Cross_Basel_1993,MELines_oxford_1977}. A common feature for ferroelectric materials is the formation of domain structures when the temperature is cooled through the ferroelectric transition temperature that is also known as the Curie Temperature \cite{Li_Actamateria_2002_subs}. For example, from a cubic to tetragonal transformation in ferroelectrics, there are six possible domains separated by the so-called domain walls, with the polarization along or opposite to the [100], [010], and [001] directions of cubic paraelectric phase \cite{Chen_JACS_2008_review}. Ferroelectric polarization switching not only depends on domain wall motion but also is influenced by the defects such as dislocations and preexisting domains as well as electrostatic field \cite{Choudhury_APL_2008_effect, Wang_Actamateria_2007_phase}. Thus, the fundamental understanding of the stability of domains and their responses to external electric field and electrostatic interactions is critical for many applications of ferroelectrics. 

In the past few decades, ferroelectric thin films have been extensively studied both theoretically and experimentally \cite{Choi_Science_2004_enhance, Haeni_Nature_2004_room, Pertsev_PRL_2000_polar, Sepliarsky_JAP_2002_ferro, Speck_JAP_1994_domain, Speck_JAP_1994_domain2,  Tenne_PhyRB_2004_abs}, owing to their many potential applications in electronic and optical devices, including data storage, sensors, non-volatile memories, thin film capacitors, etc \cite{Cross_Basel_1993,Dragan_RepProgPhy_1998,Haertling_JACS_1999_ferro}.
In traditional theoretical analysis, a particular domain wall orientation was usually given as {\it a priori} in a given domain structure, thus significantly constraining the patterns to be studied. Recently, phase field method has been successfully applied to predict the temporal domain evolution during a ferroelectric transition, offering a powerful approach to characterize the detailed domain structures in three-dimensional (3D) ferroelectric thin films without any {\it a priori} assumptions with regard to the possible domain structures \cite{Li_JAP_2004_ferro, Zhang_JAP_2008_computer}. Phase field method is able to predict not only the domain structures and the volume fractions of different orientation domains under the effect of applied external condition, such as the substrate constraint and electrostatic interactions, but also the detailed polarization switching during a ferroelectric transition \cite{Britson_Actamateria_2016_phase, Choudhury_APL_2008_effect, Li_JAP_2004_ferro,Wang_Actamateria_2007_phase, Zhang_SIAMscicomp_2009_diffuse, Zhang_JAP_2008_computer}.
The phase field model of a ferroelectric thin film is briefly reviewed in \cref{sec:2}.

When using a phase field approach to model the ferroelectric thin film phase transition,  the system usually involves both
the polarization distributions, which are the main phase field variables to depict the polarization of ferroelectric materials,
and the electrostatic potential that incorporates the electrostatic interactions. As a popular practice, their relation 
is described by the electrostatic equilibrium equation that can be derived from the Maxwell's equation.
This implies that the electrostatic field is in its equilibrium state for a given polarization field and there is no free charge inside the film. Making use of this observation, the electrostatic potential could be acquired by solving the electrostatic equilibrium equation when given a polarization distribution during the ferroelectric phase transition. In this way the electrostatic energy and the electrostatic force could also be obtained \cite{Li_JAP_2004_ferro}.

We present new variational formulations for the phase field model involving electrostatic contributions under the no free charge assumption and with periodicity in directions parallel to the film. In \cref{sec:3}, the formulations are derived for the cases involving different boundary conditions (BCs) in the direction perpendicular to the film. By expressing the energy and forces in terms of the polarization distribution only and the electrostatic interactions implicitly accounted for,  the new formulations avoid imposing additional constraints for energy minimization and temporal evolution and eliminate ambiguities that may surface in the previous formulations involving Lagrange multipliers.
 The analytically and explicitly formulated systems involve hybrid real space and Fourier space representations that are convenient to use in studies of energy landscape and relaxation dynamics.  As an illustration, we present 3D numerical simulations of phase transition in the cubic thin film of lead titanate (PbTiO$_3$) based on the new formulations in \cref{sec:4}. Some conclusions are given in \cref{sec:5}.

\section{Phase field model of a ferroelectric thin film}
\label{sec:2}
In the phase field approach, a ferroelectric domain structure in a thin film is often described by the primary order parameter $\bP(\boldsymbol{x})=(P_1,P_2,P_3)$, depicting the local spatial distribution of  polarization in the 3D space, where $\boldsymbol{x}=(x, y, z)$  being the Cartesian coordinates. The temporal evolution of the polarization vector $\bP$ is described by the time dependent Ginzburg-Landau (TDGL) equations,
\begin{equation}
\label{eq:tdgl}
\frac{\partial P_i(\boldsymbol{x},t)}{\partial t}=-\eta\frac{\delta F}{\delta P_i(\bx,t)},\quad i=1,2,3,
\end{equation}
where $F$ is the total free energy of the system and $\eta$ is the kinetic coefficient related to domain-wall mobility. $\delta F/\delta P_i(\bx,t)$ is the thermodynamic driving force for the spatial and temporal evolution of $P_i(\bx,t)$. The total free energy density includes three parts: the ferroelectric bulk free energy density $f_{bulk}(\bP)$, the domain wall energy density $f_{wall}(\bP)$, and the electrostatic energy density $f_{ele}(\bP,\bE)$. The bulk free and domain wall energy density are described respectively using the expression\cite{Li_Actamateria_2002_subs}:
\begin{equation}
\begin{aligned}
\label{eq:bulk}
f_{bulk}&(\bP)=\alpha_1(P_1^2+P_2^2+P_3^2)+\alpha_{11}(P_1^4+P_2^4+P_3^4)\\&+\alpha_{12}(P_1^2P_2^2+P_2^2P_3^2+P_1^2P_3^2)+\alpha_{111}(P_1^6+P_2^6+P_3^6)\\&+\alpha_{112}[P_1^4(P_1^2+P_2^2)+P_2^4(P_1^2+P_3^2)+P_3^4(P_1^2+P_2^2)]+\alpha_{123}(P_1^2P_2^2P_3^2),
\end{aligned}
\end{equation}
 and
\begin{equation}
\begin{aligned}
\label{eq:wall}
f_{wall}(P_{i,j})=&\frac{1}{2}G_{11}(P_{1,1}^2+P_{2,2}^2+P_{3,3}^2)+G_{12}(P_{1,1}P_{2,2}+P_{1,1}P_{3,3}+P_{2,2}P_{3,3})\\
&+\frac{1}{2}G_{44}[(P_{1,2}+P_{2,1})^2+(P_{1,3}+P_{3,1})^2+(P_{2,3}+P_{3,2})^2]\\
&+\frac{1}{2}G_{44}^\prime[(P_{1,2}-P_{2,1})^2+(P_{1,3}-P_{3,1})^2+(P_{2,3}-P_{3,2})^2],
\end{aligned}
\end{equation}
where $\alpha_1$, $\alpha_{11}$, $\alpha_{12}$, $\alpha_{111}$, $\alpha_{112}$, $\alpha_{123}$ are the Landau  expansion coefficients, $G_{11}$, $G_{12}$, $G_{44}$, $G'_{44}$ are the domain wall energy coefficients, and here a comma in the subscript stands for spatial differentiation, e.g., $P_{i,j}=\partial P_i/\partial x_j, i,j=1,2,3$, with $(x_1,x_2,x_3)$ denoting the Cartesian coordinates $(x,y,z)$ respectively. \par

In this paper, we focus on the electrostatic energy and ignore possible surface and elastic energy contributions to the free energy. These simplifications are mainly for the purpose of illustration. In fact, we have noted that some detailed calculations of the elastic energy have been provided in the literature, see the case with periodic BCs \cite{Khachaturyan_Wiely_1983}, and the case where the substrate constraints are present \cite{Li_Actamateria_2002_subs}. 
 The strategy proposed in this work is in a similar spirit to
  the micro-elasticity formulation developed in \cite{Khachaturyan_Wiely_1983} that utilized an analytical formulation based on the
  Fourier representation under the spatial periodicity assumption. For the thin film, 
  periodicities are assumed only along the film directions while the other (non-periodic) BCs are
  often necessary in the direction normal to film. Thus, we propose to adopt a
 hybrid Fourier and real space representation. Although the discussion in this work is limited to this special
 case, the extension and effective integration of hybrid formulations to more general cases
 involving additional energetic contributions can be expected and will be explored in subsequent works. 
 \par
 
To begin our technical derivations, we recall  
the Gauss's law for dielectrics: $\nabla\cdot\bD=\rho_f$. Here  $\rho_f$ is the free charge density and $\bD=\epsilon\bE+\bP$ is the electric displacement, where $\bE=-\nabla\phi$ is the electric field with the electric potential $\phi$ and $\epsilon$ is the relative permittivity \cite{Griffiths_Prentice_1999}. In the existing literature \cite{Britson_Actamateria_2016_phase, Choudhury_APL_2008_effect, Wang_Actamateria_2007_phase, Zhang_JAP_2008_computer}, the electric energy in phase field models for dielectric systems has taken various mathematical forms that correspond to two different cases of the physical systems, namely, 
$\rho_f=0$ or $\rho_f\neq 0$. 
In the presence of  free charges \cite{Britson_Actamateria_2016_phase, Li_JAP_2004_ferro}, i.e., $\rho_f\neq0$,  we suppose that the dielectric material 
brings in the free charge over time. If only the incremental free charges contribute the work to the electric energy, the energy density can be described by 
\begin{equation}
\label{eq:electric1}
  f_{ele}=-\frac{1}{2}\bE\cdot\bD=-\frac{1}{2}(\bE\cdot\bP+\epsilon|\bE|^2)\ .
\end{equation}  
 In this case, Griffiths(1999) argued that equation \cref{eq:electric1} is valid for linear dielectrics \cite{Griffiths_Prentice_1999}, i.e., $\bP=\epsilon_0\xi\bE$, where $\epsilon_0$ is the vacuum permittivity, $\xi$ is called the electric susceptibility satisfying $\epsilon=\epsilon_0(1+\xi)$. Thus, equation \cref{eq:electric1} can be written as
\begin{equation}
\label{eq:electric2}
f_{ele}=-\frac{2\epsilon-\epsilon_0}{2(\epsilon-\epsilon_0)^2}\bP^2\ ,
\end{equation} 
which is directly expressed as a function of the polarization field $\bP$. Hence, 
the electric energy  and its variation can be readily obtained, similarly as the bulk free energy and its variation. In fact,
we have
 $\dfrac{\delta F_{ele}(\bP)}{\delta\bP}=-\dfrac{2\epsilon-\epsilon_0}{(\epsilon-\epsilon_0)^2}\bP$.

Let us make a note on this case of free charges. Usually, the coefficient $\alpha_1$ of bulk free energy density in \cref{eq:bulk} has a linear temperature dependence based on Curie-Weiss law, i.e.,  $\alpha_1=\alpha(T-T_c)$, $\alpha$ is a constant, and $T_c$ is the Curie-Weiss temperature. By adding the electric energy from equation \cref{eq:electric2}, the Curie-Weiss temperature actually becomes:  $T_c^{'}=T_c+\frac{2\epsilon-\epsilon_0}{2\alpha(\epsilon-\epsilon_0)^2}$, meaning that the effect of electric field can let the polarization occur more easily
and the ferroelectric phase more stable with respect to the temperature change. \par

On the other hand, if the electric field $\bE$  or the electrostatic energy of a domain structure is considered to be self-electrostatic corresponding to the long-range electrostatic interaction of spontaneous polarizations \cite{Wang_Actamateria_2007_phase, Zhang_JAP_2008_computer}, it is natural to assume that the system satisfies the electrostatic equilibrium condition, i.e., $\rho_f=0$, thus the associated electric energy density in the case of bound charges is given by
\begin{equation}
\label{eq:electric3}
f_{ele}=-\frac{1}{2}\bE\cdot\bP\ .
\end{equation}
Moreover, based on Gauss's law, the corresponding electrostatic equilibrium condition can be written as
\begin{equation}
\label{eq:1}
\epsilon\Delta\phi=\nabla\cdot\bP\ .
\end{equation}
The equation \cref{eq:1} holds in $\Omega=(-L/2,L/2)^2\times(0,h)$ where $2L$ specifies the period along each
of the directions parallel to the film and $h$ specifies the film thickness.

In particular,  if the electric potential $\phi$ at the top and bottom surface of the film takes on constant values, we get
\begin{equation}
\label{eq:2}
\phi|_{z=0}=c_1,\quad \phi|_{z=h}=c_2, 
\end{equation}
for two constants $c_1$ and $c_2$. We call this set of condition the constant BC, particularly when $c_1=c_2$, which is named the short circuit BC.\par

We also consider the electric tip-induced BC \cite{Britson_Actamateria_2016_phase} that is defined to model the applied electric field using the Piezoresponse Force Microscopy (PFM)\cite{Choudhury_APL_2008_effect}. Then, the potential distributions on the top and bottom surface are approximated by
\begin{equation}
\label{eq:tip}
\phi|_{z=h}=\phi_{top}(x,y), \qquad \phi|_{z=0}=0, 
\end{equation}
where 
\begin{equation*}
\phi_{top}(x,y)=\phi_0\frac{\gamma^2}{(x-x_0)^2+(y-y_0)^2+\gamma^2},
\end{equation*}
here $(x_0,y_0)$ is the location of the tip, $\phi_0$ is a constant (peak of the potential), and $\gamma$ stands for the effect length scale of the potential distribution. \par

 If the normal component of the electric displacement $\bD$ is zero at that surface, i.e.,
\begin{equation}
\label{eq:3}
D_3|_{z=0}=D_3|_{z=h}=0, \;\mbox{or equivalently}, \; (\epsilon \nabla \phi - \bP)\cdot \vec{n} \mid_{z=0, h}=0, 
\end{equation}
where $\vec{n}$ is the unit vector normal to the top and bottom surface. This set of condition is named as the open circuit BC. \par

Viewing the equilibrium condition in \cref{eq:1} as an elliptic equation for the electric potential $\phi$,  we see that
conditions \cref{eq:2} and \cref{eq:tip} are Dirichlet-type BCs while \cref{eq:3} is a
Neumann-type BC. Please note that in the following calculations and derivations, the Dirichlet data $c_1$ and $c_2$ are not required to be constants unless specifically mentioned.
With the above relation between $\phi$ and polarization vector $\bP$, we can calculate the electric potential,
and then obtain the electric field, the electrostatic energy and its variation with respect to the polarization $\bP$. The details are given in \cref{app:a} and \cref{app:b}.\par

Meanwhile, we note that if one wants to consider the energy density for the system in both bound charge case and free charge case in a unified setting, a straightforward way is to use a linear combination of \cref{eq:electric1} and \cref{eq:electric3}, i.e., 
$$
f_{ele}=-(\beta\bE\cdot\bP+(1-\beta)\bE\cdot\bD)\ ,
$$
for a constant $\beta\in[0,1]$. For the special case where $\beta=0$ or $\beta=1$, we have the specified relation between $\bE$ and $\bP$ described above respectively. However, for other values of $\beta$, it is unclear which specific relation between electric potential and polarization vector remains applicable. 

\section{New variational formulations of electrostatic interactions}
\label{sec:3}

We now focus on the case of bound charges with the electric energy density given by equation \cref{eq:electric3} and subject
to the electrostatic equilibrium condition \cref{eq:1}.  We note that
an explicit solution to equation \cref{eq:1} can be used to not only simplify the phase field energy formulation \cref{eq:electric3}, but also derive an explicit mathematical expression of the functional variation of the electrostatic energy.  This 
effectively allows us to 
find the explicit mathematical expression of the total driving force: \par
 \begin{equation}
 \frac{\delta F}{\delta P_i}=\frac{\delta F_{bulk}}{\delta P_i}+\frac{\delta F_{wall}}{\delta P_i}+\frac{\delta F_{ele}}{\delta P_i},\qquad i=1,2,3\,,
 \end{equation}
in terms of the polarization vector $\bP$.
 Here we use $ (x_1,x_2,x_3)$ to denote Cartesian coordinates $(x,y,z) $ respectively. First of all, 
 the bulk and the domain wall driving forces are given respectively by
\begin{equation}
\begin{aligned}
\frac{\delta F_{bulk}}{\delta P_i}=&2\alpha_1P_i+4\alpha_{11}P_i^3+2\alpha_{12}P_i\sum_{j\neq i}P_j^2
+6\alpha_{111}P_i^5 +4\alpha_{112}P_i^3\sum_{j\neq i}P_j^2\\
&+4\alpha_{112}P_i\sum_{j\neq i}P_j^4+2\alpha_{123}P_i\prod_{j\neq i}P_j^2, \qquad   i,j=1,2,3\ , 
\end{aligned}
\end{equation}
and
\begin{equation}
\begin{aligned}
\frac{\delta F_{wall}}{\delta P_i}=&-G_{11}\frac{\partial^2 P_i}{\partial x_i^2}-(G_{44}+G'_{44})\sum_{j\neq i}\frac{\partial^2 P_i}{\partial x_j^2}\\
&+(G'_{44}-G_{12}-G_{44})\sum_{j\neq i}\frac{\partial^2P_j}{\partial x_i\partial x_j}, \qquad \qquad i,j=1,2,3.
\end{aligned}
\end{equation}
The electrostatic driving forces, i.e., the variations of the energy given in \cref{eq:electric3}, subject to the 
 electrostatic equilibrium condition \cref{eq:1} and
 various BCs, are given by
\begin{equation}
\label{eq:tab1}
\frac{\delta F_{ele}}{\delta\bP}=
\left\{
\begin{array}{ll}
\nabla\phi-\frac{1}{2}\nabla\phi_2\, , \quad   &\mbox{Dirichlet BC, e.g., constant BC \cref{eq:2},  tip BC \cref{eq:tip}, }\\
 \nabla\phi \, , &\mbox{Neumann BC, e.g., open circuit BC \cref{eq:3},}
 \end{array}
\right.
\end{equation}
where $\phi_2$ is an auxiliary potential. 
While more detailed derivations are given in \cref{app:b}, we offer the main procedures on how the terms ($\nabla \phi$ and $\nabla \phi_2$) in the \cref{eq:tab1} are determined.
 Let us use $\hat{(\cdot)}$ to denote the 2D Fourier series expansion due to the periodicity of  $\phi$ and $\bP$ in the $x_1 $-$x_2$ plane with $(\lambda_1,\lambda_2)$ being the variables in the Fourier (frequency) space.\par
 
Let us work with $\phi_2$ first for the constant and tip-induced boundary cases with more general discussions given in \cref{app:a}.  For the constant BC of equation \cref{eq:2}, when $c_1$ and $c_2$ are all constants, we can take $\nabla\phi_2=\ba$ where $\ba=(0,0,a)^{T}$, 
$a=\frac{c_2-c_1}{h}$ with $h$ being the film thickness.\par
 As for the tip-induced BC of equation \cref{eq:tip}, based on the detailed calculation of $\phi_2$ given in \cref{app:tip}, the function $\phi_2$ is recovered from its Fourier representation 
given by
\begin{equation}
\hat{\phi}_2(\lambda_1,\lambda_2,z)=\frac{\hat{\phi}_{top}(\lambda_1,\lambda_2)
}{M(h)}(e^{|\mathbf{\lambda}| z}-e^{-|\mathbf{\lambda}| z}),
\end{equation}
where $\hat{\phi}_{top}$ is the Fourier expansion of the potential  $\phi|_{z=h}=\phi_{top}$ on the top surface, 
 $|\mathbf{\lambda}|=\sqrt{\lambda_1^2+\lambda_2^2}$ and
 $M(h)= e^{|\mathbf{\lambda}| h}-e^{-|\mathbf{\lambda}| h}$. \par
 
Next, we condition the determination of $\phi$ when $(\lambda_1,\lambda_2)\neq (0,0)$, and equation \cref{eq:1} leads to
\begin{equation}
\label{eq:solve}
\epsilon(\frac{\partial^2\hat{\phi}}{\partial z^2}-\lambda_1^2\hat{\phi}-\lambda_2^2\hat{\phi})=I\lambda_1\hat{P_1}+I\lambda_2\hat{P_2}+\frac{\partial\hat{P_3}}{\partial z}\triangleq f(\lambda_1,\lambda_2,z), 
\end{equation}
where $I=\sqrt{-1}$ and $f=f(\lambda_1,\lambda_2, z)$ denotes the 2D Fourier representation of the divergence 
on the right hand side in equation \cref{eq:1}.  The function $\phi$ is recovered from its Fourier representation given by
\begin{equation}
\label{eq:solve1}
\hat{\phi} (\lambda_1,\lambda_2, z)=C_1(\lambda_1,\lambda_2)e^{|\mathbf{\lambda}| z}+C_2(\lambda_1,\lambda_2)e^{-|\mathbf{\lambda}| z}+g(\lambda_1,\lambda_2, z),
\end{equation}
where the function $g=g(\lambda_1,\lambda_2,z)$ is defined by 
\begin{equation}
\label{eq:gz}
g(\lambda_1,\lambda_2,z)
= \frac{1}{2|\mathbf{\lambda}|\epsilon}\int_0^{z} [ (e^{|\lambda|(z-s)}-e^{-|\lambda|(z-s)})  f(\lambda_1,\lambda_2,s)  ]\ ds.
 \end{equation}
As for the coefficients $\boldsymbol{C}=(C_1,C_2)^\intercal$ under the Dirichlet BC with $c_2$ and $c_1$  being the top and bottom boundary data respectively, we have
\begin{equation}
\label{eq:Coe2}
 \boldsymbol{C}(\lambda_1,\lambda_2)=\frac{1}{ M(h) }\begin{pmatrix}
 -\hat{c}_1(\lambda_1,\lambda_2)e^{-|\lambda|h}+\hat{c}_2(\lambda_1,\lambda_2)-g(\lambda_1,\lambda_2,h) \\
 \hat{c}_1(\lambda_1,\lambda_2)e^{|\lambda|h}-\hat{c}_2(\lambda_1,\lambda_2)+g(\lambda_1,\lambda_2,h)
\end{pmatrix}, 
\end{equation}
where $\hat{c}_1$ and $\hat{c}_2$ are the Fourier representations of  $c_1$ and $c_2$. \par

For the constant BC \cref{eq:2} with $c_1$ and $c_2$ being constants, we have
 $\hat{c}_1(\lambda_1,\lambda_2)=\hat{c}_2(\lambda_1,\lambda_2)=0$.
So, equation \cref{eq:Coe2} is effectively given by $C_2(\lambda_1,\lambda_2)= -C_1(\lambda_1,\lambda_2)=g(\lambda_1,\lambda_2,h)/M(h)$.

Meanwhile, under the tip-induced condition \cref{eq:tip},  we get instead
\begin{equation}
\label{eq:Coetip}
\boldsymbol{C}(\lambda_1,\lambda_2)= \frac{1}{ M(h) } 
\begin{pmatrix}
\hat{\phi}_{top}(\lambda_1,\lambda_2)-g(\lambda_1,\lambda_2,h)\\
-\hat{\phi}_{top}(\lambda_1,\lambda_2)+g(\lambda_1,\lambda_2,h)
\end{pmatrix}.
\end{equation} 
Finally, under the open circuit BC \cref{eq:3}, we have 
\begin{equation}
\label{eq:Coe3}
\boldsymbol{C}(\lambda_1,\lambda_2)=\frac{1}{ |\lambda| M(h) } \begin{pmatrix}
 -\hat{P}_3(\lambda_1,\lambda_2,0)e^{-|\lambda|h}+\hat{P}_3(\lambda_1,\lambda_2,h)- g_3(\lambda_1,\lambda_2,h) \\
 -\hat{P}_3(\lambda_1,\lambda_2,0)e^{|\lambda|h}+\hat{P}_3(\lambda_1,\lambda_2,h)-g_3(\lambda_1,\lambda_2,h)
\end{pmatrix},
\end{equation} 
where
\begin{equation}
\label{eq:gzd} 
g_3(\lambda_1,\lambda_2,h)=\frac{1}{2\epsilon}\int_0^{z}
[(e^{|\lambda|(z-s)}+e^{-|\lambda|(z-s)})f(\lambda_1,\lambda_2,s)] \ ds
\end{equation} 
denotes the partial derivative of $g(\lambda_1,\lambda_2,z)$ with respect to the third variable $z$. \par

It is important to highlight that as $f=f(\lambda_1,\lambda_2, z)$ is solely computed from the polarization field $\bP$,
so are the functions  $g$, $g_3$ and  $\hat{\phi} $.
The numerical computations of the integrals associated with both $g$ and $g_3$  are highly dependent on the discretization 
of the functions and differential equations along the $z$ direction.  For illustration, here we adopt a finite difference approximation in the $z$ direction on a uniform grid, which allows us to 
conveniently apply the composite Simpson's rule based on the same grid points without further interpolations.
We leave more detailed analysis of numerical discretization in subsequent works.

As a final note added for implementing the new variational formulations, we remark that for $\lambda_1=\lambda_2=0$, \cref{eq:solve} should be modified. 
For this special case, equation \cref{eq:solve} can be simplified as a second order ordinary differential equation for the
 real variable $z$ with the solution given by
 $$\hat{\phi}(0,0,z)=\int_0^z \hat{P}_3(0,0,s)ds+B_1 z+B_2$$
  where the $B_1$ and $B_2$ are determined by different BCs, e.g., 
 $$
 \left\{
 \begin{array}{lll}
 \displaystyle
  B_1=\frac{1}{h} \left(\hat{c}_2-\hat{c}_1-\int_0^h \hat{P}_3(0,0,s)ds\right),\quad 
 B_2=\hat{c}_1,\quad &  \mbox{  for condition \cref{eq:2},}\\ 
 \displaystyle
 B_1=\frac{1}{h} \left(\hat{\phi}_{top}(0,0)-\int_0^h \hat{P}_3(0,0,s)ds\right),\quad
B_2=0,\quad &  \mbox{ for condition \cref{eq:tip},}\\
  B_1=B_2=0, \quad & \mbox{ for condition \cref{eq:3}.}
\end{array}\right.
$$ \par  

With the derivations above, the equation \cref{eq:tdgl} can be numerically solved 
by various methods. Although we leave detailed discussion on the numerical approximations to separate works, illustrative
examples are presented later to show the effectiveness of the new formulations here.

\section{Illustrative examples}
\label{sec:4}

We now present two numerical examples of the new variational formulations of the total energy and the driving forces by computing the equilibria and transition states.
We adopt the steepest descent gradient dynamics (a.k.a., TDGL) for the former, which leads to the equilibria via temporal polarization evolution. 
For the latter, the transition state is the point (state) of the highest energy along the minimum energy path (MEP) between two local equilibria, which characterizes the morphology of critical nucleus and the critical nucleation energy that determines the rate of a nucleation reaction \cite{Zhang_PRL_2007_morph, Zhang_JCP_2010_diffuse, Zhang_npj_2016_recent}.
In the last few decades, various numerical methods have been developed for saddle points and MEPs calculation, e.g., the dimer method and its improvements \cite{Levitt_SIAMnumer_2017, Zhang_SIAMnumer_2012_shrink, Zhang_SIAMJSCom_2016_opt}, the Nudged Elastic Band (NEB) method \cite{Henkelman_JChemP_2000_a}, and the string method as well as its various improvements \cite{Du_ComMatSci_2009_a,E_PhyRB_2002_string,E_JChemP_2007_simplify}. In this paper, we adopt a simplified string method in \cite{E_JChemP_2007_simplify} to calculate the MEP to show the complete $180^{\circ}$ polarization switching process described by the MEP connecting two equilibria.

\subsection{Numerical results}
We take the lead titanate (PbTiO$_3$) thin film as an example. The simulations are done on 3D computational domain of the size  $64\Delta x\times64\Delta x\times64\Delta x$ with the parameter  $\Delta x=1.0$ nm referring to a uniform grid spacing in all three coordinate directions.
The coefficients of the bulk free energy are exactly taken from \cite{Li_Actamateria_2002_subs}. Here the vacuum permittivity  $\epsilon_0=8.85\times10^{-12}Fm^{-1}$, and the electric susceptibility $\epsilon=100\epsilon_0$. The isotropic domain wall energy coefficients are taken to be $G_{11}/G_{110}=\frac{1}{2\Delta x}$, $G_{12}/G_{110}=0$, and $G_{44}/G_{110}=G'_{44}/G_{110}=\frac{1}{4\Delta x}$, where $G_{110}$ is related to the magnitude of grid spacing $\Delta x$ via $\Delta x=\sqrt{G_{110}/\alpha_0}$ and $\alpha_0=1.7252\times10^8C^{-1}m^2N$. \par
To simulate the temporal polarization change of the domain wall or the phase transition of ferroelectric in the presence of electric filed, the relaxation system equation \cref{eq:tdgl} is solved by using the semi-implicit Fourier spectral method with periodic BCs in $x_1$ and $x_2$ axis along the film plane \cite{Chen_ComPhyComm_1998_apply}. We compute several phase transitions by applying our newly derived electrostatic energy variation with different BCs in \cref{eq:tab1}. In the figures presented here, different colors (red and blue) are used to represent the equivalent polarization magnitude and the corresponding polarization direction, i.e. $\bP=(0,0,1)$ and $\bP=(0,0,-1)$, respectively. The gradual change from the blue color to the red color represents the local dipole polarization magnitude and change in direction from $-1$ to $1$, or vice versa. \par
\begin{figure}[htbp]
\centering
\includegraphics{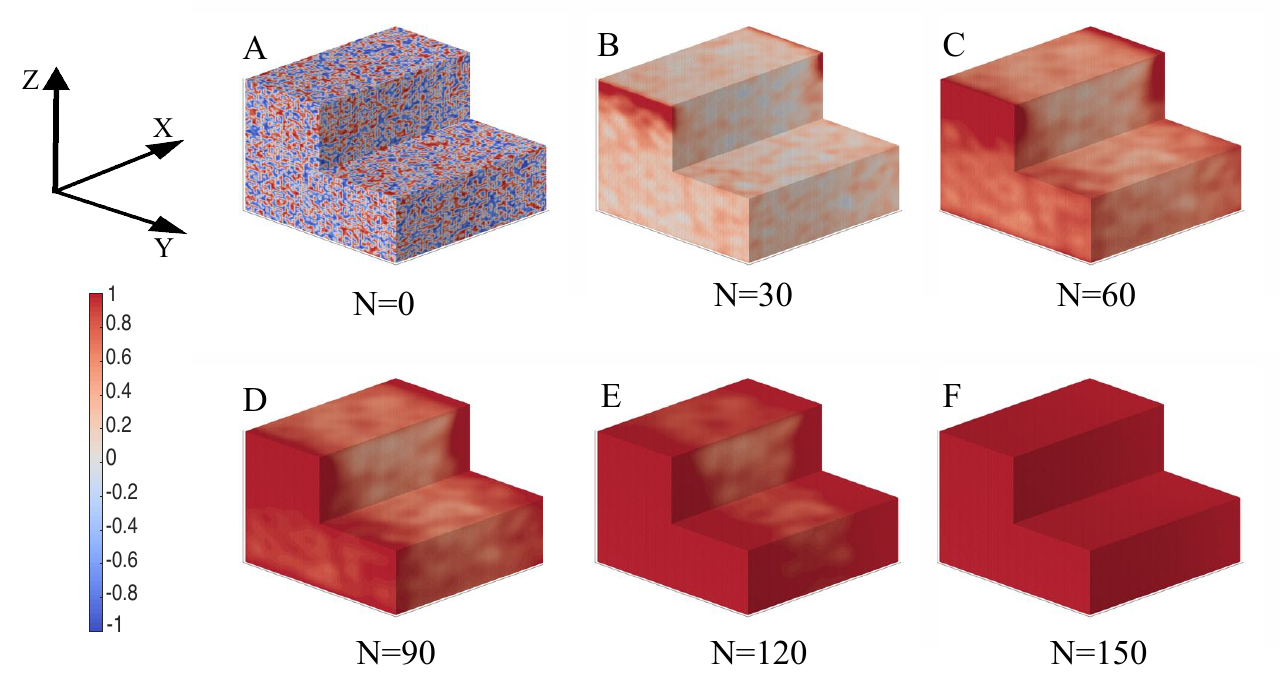}
\caption{\textbf{Ferroelectric phase transition from an initial random polarization distribution to an equilibrium under the constant BC.} A-F: the simulated domain wall changes in the 3D configuration space at different iteration steps, e.g., N=0:30:150, respectively.}
\label{fig:1}
\end{figure}
\begin{figure}[htbp]
\centering
\includegraphics{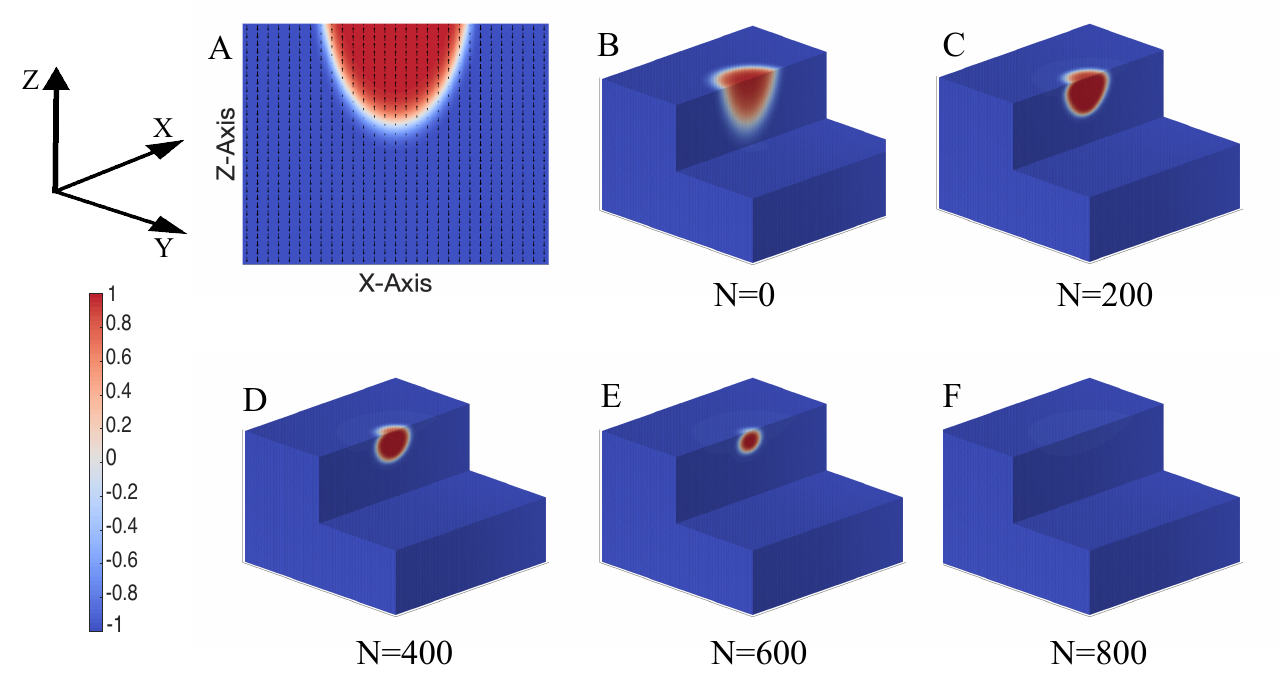}
\caption{\textbf{Ferroelectric phase transition from an initial polarization distribution to an equilibrium under the tip-induced BC.} A: the sliced view (at $y=0$) of the initial polarization state of the ferroelectric domain wall in X-Z plane (blue and red-shade plane in the schematic). B-F: the simulated domain wall changes in the 3D configuration space at different iteration steps, e.g., N=0:200:800, respectively.}
\label{fig:2}
\end{figure}

\begin{figure}[htbp]
\centering
\includegraphics{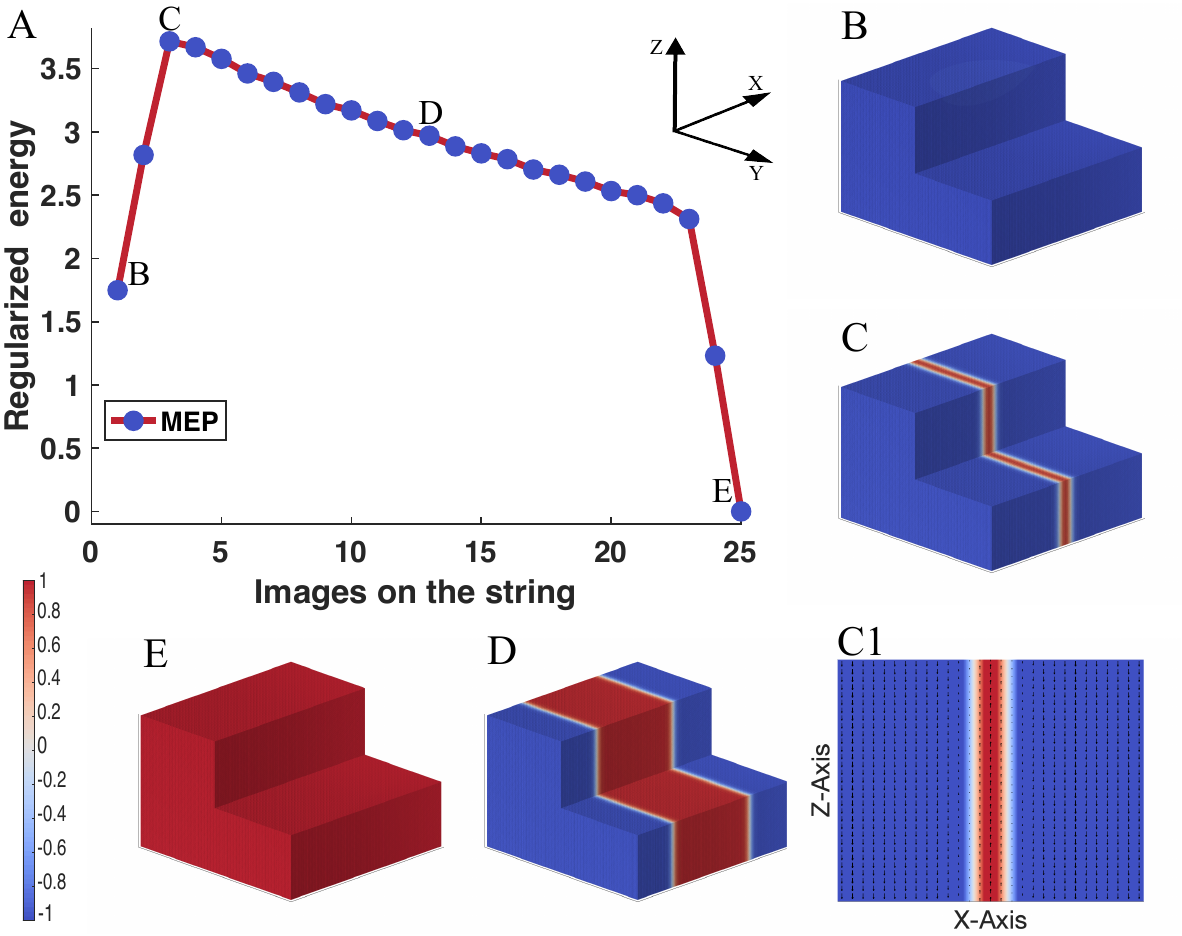}
\caption{\textbf{ Computed MEP shows the ribbon-like $180^\circ$ polarization switching process under the constant BC.}  A: the MEP of polarization switching connecting the initial polarization state (B) with the final polarization state(E) by passing through the critical nucleus (C).  B: the initial steady polarization state ($\bP=(0,0,-1)$); C: a ribbon-like critical nucleus corresponding to the point with the highest energy on the MEP; C1:  sliced view of the critical nucleus (at $y=0$) showing the domain pattern in the X-Z plane; D: an intermediate state after passing the critical nucleus, corresponding to the point D on  the MEP; E: the final steady polarization state ($\bP=(0,0,1)$).}
\label{fig:3}
\end{figure}

\Cref{fig:1} shows the numerical simulation of the ferroelectric phase transition started from a random domain distribution to an equilibrium in the presence of electric field. 
In the example, the electric field in the form of \cref{eq:electric3} is used with the constant BC \cref{eq:2}, in which the electric potential on the top surface of the film is lower than that on the bottom surface. From the result, the random domains gradually disappear under a large enough electric potential during the evolution, and the final polarization domain is formed to minimize the electric effect.\par
\Cref{fig:2} shows the ferroelectric phase transition started from a tip-induced-like domain configuration to an equilibrium in the presence of electric filed. \cref{fig:2}A shows the sliced view (at $y=0$) of the initial polarization state of \cref{fig:2}B in X-Z plane, and the scale and direction of black arrows illustrate the local dipole magnitude and direction of each unit cell in X-Z plane. In this case, the electric field is in the form of equation \cref{eq:electric3} with the tip-induced BC \cref{eq:tip}, and the tip-induced electric potential is negative on the top surface of the film, but  zero on the bottom surface of the film. While the electric field effect may generate a local polarization switching, numerical simulation shows that it cannot induce a complete polarization switching
  if the initial domain (nucleus) is not large enough or the electric filed is not strong enough. \par
Next, we apply the simplified string method to compute the complete process of $180^\circ$ polarization switching from $\bP=(0,0,-1)$ to $\bP=(0,0,1)$ with the new variational form. In \cref{fig:3}, we plot the MEP of total polarization switching process, which corresponds to a ribbon-like pattern geometrically. In the presence of electric field, \cref{fig:3}C shows the configuration of critical nucleus as a thin polarization switching domain with sharp interface, which gives the width of interface $\delta\approx\sqrt{G_{11}/\alpha_0}$, and $\delta=\sqrt{\Delta x/2}$ in this case. We slice the configuration of nucleus along $y=0$ axis to show its domain pattern in 2D X-Z plane clearly, with the scale and direction of black arrows indicating the local dipole magnitude and direction of each unit cell in X-Z plane. Moreover, the width of the thin domain gets enlarged with the increase of energy, and once the critical nucleus is formed to overcome the energy barrier, the switched polarization domain continues to grow until the final equilibrium state $\bP=(0,0,1)$ is achieved.\par

\section{Conclusions/Summary}
\label{sec:5}
With the newly formulated phase field energy involving electrostatic energy contributions, the 
variation of the electrostatic energy and the total free energy becomes straightforward. It eliminates
the requirement to impose constraints in the variational calculation and avoids the use of associated Lagrange multipliers. The explicitly formulated expression of the driving force makes it convenient for numerical simulations and avoids ambiguity. It also helps to improve simulation accuracy as the electrostatic equilibrium equation is now given by an exact explicit analytical solution.\par

By using the new electrostatic energy variation under different BCs with suitable electrostatic potential, we are able to accurately perform the complete phase transition process of the $180^{\circ}$ polarization switching and find the critical nucleus with a thin polarization domain in 3D  configuration space while demonstrating the effectiveness of the mathematical formulation. The shapes of critical nuclei could be varied under different electric fields and driving forces. More detailed discussions on other types of polarization switching process will be illustrated in a later work. \par

Although the current work only focuses on the ferroelectric phase transition involving electrostatic contributions, the other contributions such as the elastic energy can be also taken into account.
We expect that similar approach can be applied to more general phase field models for ferroelectric and ferromagnetic materials,
which will be pursued in future.

\appendix

\section{Calculating the electric field under different BCs }
\label{app:a}
We have already presented the precise functional forms of the electrostatic energy given respectively in the equation \cref{eq:electric1} and \cref{eq:electric3}. We now present in more details on calculating the solution of the electrostatic equilibrium equation \cref{eq:1} subject to either \cref{eq:2} or \cref{eq:3}, and the functional variation of the energy form \cref{eq:electric3} under different BCs. 

We first discuss how to calculate the electric field and the electric potential $\phi$ from the polarization field.
Applying the 2D Fourier expansion on the equation \cref{eq:1}, as the $\phi$ and $\bP$ are periodic in the X-Y (or equivalently the $x_1$-$x_2$) plane,  equation \cref{eq:1} becomes \cref{eq:solve}, i.e.,
\begin{equation*}
\hat{\phi}(\lambda_1,\lambda_2,z)=\frac{1}{L^2}\int_{-\frac{L}{2}}^{\frac{L}{2}}\int_{-\frac{L}{2}}^{\frac{L}{2}}\phi(x,y,z)e^{-I(\frac{2\pi\lambda_1x+2\pi\lambda_2y}{L})}\ dxdy\,.
\end{equation*}
For each $\lambda_1$ and $\lambda_2$,  equation \cref{eq:solve} can be taken as an independent scalar linear ordinary differential equation of $\hat{\phi}$ in real variable $z$, and we can easily find its general solution in the form of equation \cref{eq:solve1}.

By performing a 2D Fourier expansion on the electric BC, equation \cref{eq:2}, \cref{eq:3} and \cref{eq:tip}, the unknown coefficients $C_1=C_1(\lambda_1,\lambda_2)$ and $C_2=C_2(\lambda_1,\lambda_2)$ can be determined respectively. More specifically, let $g(z)$ and $g_3(z)$  be short-hand notations of the functions 
$g(\lambda_1,\lambda_2, z)$ and $g(\lambda_1,\lambda_2, z)$ given by equation \cref{eq:gz} and 
\cref{eq:gzd} respectively, with $f(\lambda_1,\lambda_2, z)$ being defined 
by \cref{eq:solve},  we have the following 2$\times$2 linear systems. \par
For the Dirichlet BC (including the special constant BC by equation \cref{eq:2}),
\begin{equation}
\left\{
\begin{aligned}
 &C_1+C_2=\hat{c}_1(\lambda_1,\lambda_2), \\
 C_1 e^{|\mathbf{\lambda}| h}&+C_2 e^{-|\mathbf{\lambda}| h}+g(h)=\hat{c}_2(\lambda_1,\lambda_2).
\end{aligned}
\right.
\end{equation}
For the  tip-induced BC by equation \cref{eq:tip}, we have a special case of the above, namely,
\begin{equation}
\left\{
\begin{aligned}
 &C_1+C_2=0 , \\
 C_1 e^{|\mathbf{\lambda}| h}&+C_2e^{-|\mathbf{\lambda}| h}+g(h)=\hat{\phi}_{top}(\lambda_1,\lambda_2).
\end{aligned}
\right.
\end{equation}
For the open circuit (Neumann) BC by equation \cref{eq:3}, we have
\begin{equation}
\label{eq:open}
\left\{
\begin{aligned}
 &C_1|\mathbf{\lambda}|-C_2|\mathbf{\lambda}|=\hat{P_3}(\lambda_1,\lambda_2,0), \\
 2|\mathbf{\lambda}|(C_1e^{|\mathbf{\lambda}| h}&-C_2e^{-|\mathbf{\lambda}| h})+g_3(h) =\hat{P_3}(\lambda_1,\lambda_2,h),
\end{aligned}
\right.
\end{equation}
where $g_3(h)$ is the value of  $g_3(z)$  at $z$ equal to $h$.
By solving these linear equations, we can get the coefficients $C_1$ and $C_2$ corresponding to different BCs, as given in equation \cref{eq:Coe2}, \cref{eq:Coetip} and \cref{eq:Coe3} respectively.
 The electric potential and electric field can be easily obtained from the Fourier expansion (or the discrete inverse Fourier transform
 on lattice points) of equation \cref{eq:solve1} and its derivatives, e.g.,
\begin{equation}
\label{eq:invphi}
\phi(x,y,z)={L^2} \sum_{\lambda_1,\lambda_2=-\infty}^{\infty}
\hat{\phi}(\lambda_1,\lambda_2,z)e^{I\frac{2\pi\lambda_1x}{L}}e^{I\frac{2\pi\lambda_2y}{L}}
\end{equation}
and 
\begin{equation}
\label{eq:invphi1}
\left( \begin{array}{c}
\frac{\partial\phi}{\partial x}\\
\frac{\partial\phi}{\partial y}
\end{array}\right)
={L^2}\sum_{\lambda_1,\lambda_2=-\infty}^{\infty}I
\left(\begin{array}{c} \lambda_1\\
\lambda_2
\end{array}\right)
\hat{\phi}(\lambda_1,\lambda_2,z)e^{I\frac{2\pi\lambda_1x}{L}}e^{I\frac{2\pi\lambda_2y}{L}},
\end{equation}
while
\begin{equation}
\frac{\partial\phi}{\partial z}=L^2\sum_{\lambda_1,\lambda_2=-\infty}^{\infty}[|\mathbf{\lambda}|(C_1e^{|\mathbf{\lambda}| z}-C_2e^{-|\mathbf{\lambda}| z})+g_3(z)]e^{I\frac{2\pi\lambda_1x}{L}}e^{I\frac{2\pi\lambda_2y}{L}}.
\end{equation}
Consequently, the electrostatic energy can also be obtained.

For the Poisson equation \cref{eq:1} with a more general Dirichlet BC in the $z$ direction, we can make use of the linearity of the operator $\Delta$, and decompose the solution $\phi$ of equation \cref{eq:1} into $\phi=\phi_1+\phi_2$ with $\phi_1$\ , $\phi_2$\ satisfying
\begin{equation}
\label{eq:diri}
\left\{
\begin{aligned}
&\epsilon\Delta\phi_1=\nabla\cdot\bP\\  
&\phi_1|_{z=0}=\phi_1|_{z=h}=0 
\end{aligned}\right.
\qquad \mbox{and} \qquad 
\left\{
\begin{aligned}
&\epsilon\Delta\phi_2=0\\  
&\phi_2|_{z=0}=c_1,\quad \phi_2|_{z=h}=c_2. 
\end{aligned}\right.
\end{equation}
The total solution for equation \cref{eq:1} under this type of BC is $\phi=\phi_1+\phi_2$, and the total electric field is $\nabla\phi=\nabla\phi_1+\nabla\phi_2$. This works in general, without requiring
the Dirichlet data $c_1$ and $c_2$ being constants. 
\par

Now, we adopt the same idea and notations to calculate $\phi_2$ and its derivatives under the Dirichlet BC. For the special case by equation \cref{eq:2}, when both $c_1$ and $c_2$ are constants,
 the solution $\phi_2$ with these boundary constants has the form of $\phi_2=az+b$ ($a=\dfrac{c_2-c_1}{h}$) and $b=c_1$. Thus, the special case leads to $\phi=\phi_1+az+b$,  $\nabla\phi_2=\ba$ and $\nabla\phi=\nabla\phi_1+\ba$, where $\ba=(0,0,a)^\mathrm{T}$. \par
  In general, we may
apply the 2D Fourier expansion in X-Y plane to solve the Laplace's equation for $\phi_2$, i.e., 
\begin{equation}
\label{eq:2df}
\frac{d^2\hat{\phi}_2}{dz^2}-(\lambda_1^2+\lambda_2^2)\hat{\phi}_2=0.
\end{equation}
The solution of the homogeneous second order ordinary differential equation with respect to real variable z by equation \cref{eq:2df} has the form
\begin{equation}
\hat{\phi}_2(\lambda_1,\lambda_2,z)=C_1(\lambda_1,\lambda_2)e^{|\mathbf{\lambda}| z}+C_2(\lambda_1,\lambda_2)e^{-|\mathbf{\lambda}| z},
\end{equation}
where $|\mathbf{\lambda}|$ is the same used as above. Performing a 2D Fourier expansion on the BC \cref{eq:2}, the unknown coefficients $C_1$ and $C_2$ can be determined by solving the linear equations
\begin{equation}
\left\{
\begin{aligned}
 &C_1(\lambda_1,\lambda_2)+C_2(\lambda_1,\lambda_2)=\hat{c}_1, \\
 &C_1(\lambda_1,\lambda_2)e^{|\mathbf{\lambda}| h}+C_2(\lambda_1,\lambda_2)e^{-|\mathbf{\lambda}| h}=\hat{c}_2.
\end{aligned}
\right.
\end{equation}
Thus, the solution of $C_1$ and $C_2$ could be given by
\begin{equation}
\label{eq:C12}
C_1(\lambda_1,\lambda_2)=\frac{\hat{c}_2-\hat{c}_1e^{-|\lambda|h}}{e^{|\mathbf{\lambda}| h}-e^{-|\mathbf{\lambda}| h}}, \qquad  C_2(\lambda_1,\lambda_2)=\frac{\hat{c}_1e^{|\lambda|h}-\hat{c}_2}{e^{|\mathbf{\lambda}| h}-e^{-|\mathbf{\lambda}| h}} .
\end{equation}

\section{Calculation of the electrostatic energy variation of \cref{eq:electric3} under different BCs}
\label{app:b}

\subsection{Energy and energy variation under the Dirichlet BC}
\label{sec:b2}
To calculate the electrostatic energy in the form of \cref{eq:electric3} and its energy variation under the Dirichlet BC, we adopt the splitting idea above, and the electrostatic energy in the form \cref{eq:electric3} can be rewritten as
\begin{equation*}
F_{ele} =\int_{\Omega}f_{ele} \ dV=\int_{\Omega}\frac{1}{2}(\nabla\phi_1\cdot\bP+\nabla\phi_2\cdot\bP)\ dV.
\end{equation*}
Thus, we only need to handle the term $\nabla\phi_1$, as the $\phi_2$ is independent of $\bP$ according to \cref{eq:diri}. For notation convenience, we just use $\phi_1(\bP)$ to denote the dependence of $\phi_1$ on $P$, which is a linear operator and can be easily found by the left part of equation \cref{eq:diri}. More specific,
$\phi_1(\bP)$ refers to the solution of the Poisson equation subject to the homogeneous Dirichlet boundary for a given $\bP$, i.e., $\Delta\phi_1(\bP)=\frac{1}{\epsilon}div(\bP)$ in the domain and $\phi_1=0$ on the top and bottom  boundary surface (periodicity in the film plane). Meanwhile, $\nabla\phi_1(\bP)$ refers to the electric field corresponding to the auxiliary potential $\phi_1$, and we calculate the $Fr\acute{e}chet$ derivative of electrostatic energy at $\bP$ along the direction $\vec{Q}$ as
\begin{equation}
\label{eq:frechet1}
(\frac{\partial F_{ele} }{\delta \bP}, \vec{Q})= \frac{1}{2} \int_{\Omega} [\nabla \phi_1(\bP)\cdot\vec{Q}+ \nabla \phi_1(\vec{Q}) \cdot\bP]\ dV\,.
\end{equation}
Using the divergence theorem, Green's identity and the periodicity 
of $\bP$, $\vec{Q}$ and  $\phi_1(\bP)$, $\phi_1(\vec{Q})$ in the X-Y plane
and the zero Dirichlet condition for the latter pair on the top and bottom surface, 
 we get
\begin{equation}
\label{eq:lq}
\begin{aligned}
 \int_{\Omega}\nabla \phi_1(\vec{Q})\cdot\bP dV &= -  \int_{\Omega} \phi_1(\vec{Q}) \ div\bP dV\\
  &= -  \epsilon \int_{\Omega} \phi_1(\vec{Q}) \ \Delta \phi_1(P) dV\\
  &= -  \epsilon \int_{\Omega} \Delta \phi_1(\vec{Q}) \  \phi_1(P) dV\\
   &= -  \int_{\Omega} div \vec{Q} \  \phi_1(P) dV\\
     &=   \int_{\Omega}  \vec{Q} \cdot \nabla \phi_1(P) dV\\
\end{aligned}
\end{equation}

Combing the above derivation and results in equation \cref{eq:frechet1},  we can get the variation of electrostatic energy in the form \cref{eq:electric3}, under the Dirichlet BC \cref{eq:2}, as
\begin{equation}
\label{eq:elecphi}
\frac{\delta F_{ele}}{\delta\bP}=\nabla\phi_1+\frac{1}{2}\nabla\phi_2=\nabla\phi-\frac{1}{2}\nabla\phi_2.
\end{equation}

\subsubsection{The case with constant BC}
\label{sec:b3}
A particular case of the Dirichlet BC is the constant BC by
equation \cref{eq:2}, with $c_1$ and $c_2$ being constants in X-Y plane. 
For equation \cref{eq:electric3}, we end up with the following form for the electrostatic energy
\begin{equation*}
F_{ele} =\int_{\Omega}\frac{1}{2}(\nabla\phi_1\cdot\bP+\ba \cdot \bP)\ dV,
\end{equation*}
and
the energy variation
$$
\frac{\delta F_{ele}}{\delta\bP}=\nabla\phi-\frac{1}{2}\nabla\phi_2=\nabla\phi - \frac{1}{2}\ba\,.
$$

\subsubsection{The case with tip-induced BC}
\label{app:tip}
The tip-induced BC by equation \cref{eq:tip} is another special case of the Dirichlet BC.
Following the results presented in \cref{app:a} above
and with
$\phi=\phi_1+\phi_2$,
we easily get that the electrostatic energy variation with respect to polarization vector $\bP$ in the energy density form, equation \cref{eq:electric3}, is 
$$\frac{\delta F_{ele}}{\delta\bP}=\nabla\phi_1+\frac{1}{2}\nabla\phi_2=\nabla\phi-\frac{1}{2}\nabla\phi_2\,.$$
The potential $\phi_2$ satisfies 
\begin{equation}
\label{eq:phi2}
\left\{
\begin{aligned}
&\Delta\phi_2=0\\
\phi_2|_{z=0}=0,\quad \phi_2|_{z=h}=\phi_0&[\frac{\gamma^2}{(x-x_0)^2+(y-y_0)^2+\gamma^2}].
\end{aligned}
\right.
\end{equation}
The solution can be determined as discussed previously in \cref{app:a}.
In this special case, equation \cref{eq:C12} leads to
$$
C_1(\lambda_1,\lambda_2)=-C_2(\lambda_1,\lambda_2)=\frac{\hat{\phi}_{top}(\lambda_1,\lambda_2)}{e^{|\mathbf{\lambda}| h}-e^{-|\mathbf{\lambda}| h}}.
$$
Similarly, we can get the solution of $\phi_2$ in both real and Fourier variables under the Dirichlet BC through the 2D Fourier expansion, i.e., equation \cref{eq:invphi}, and in this way the gradient of $\phi_2$ can also be computed, e.g.,  through equation \cref{eq:invphi1}, and
\begin{equation}
\frac{\partial\phi_2}{\partial z}=L^2\sum_{\lambda_1=-\infty}^{\infty}\sum_{\lambda_2=-\infty}^{\infty}|\mathbf{\lambda}|(C_1e^{|\mathbf{\lambda}| z}-C_2e^{-|\mathbf{\lambda}| z})e^{I\frac{2\pi\lambda_1x}{L}}e^{I\frac{2\pi\lambda_2y}{L}}.
\end{equation}
Thus the total electrostatic potential can be obtained by summing up the solution of the two parts in equation \cref{eq:diri}, so as the electric field and the electrostatic energy.

\subsection{Electrostatic energy and its variation in an open circuit}

We note that the electrostatic energy in equation \cref{eq:electric3} is
$$
F_{ele}=\int_{\Omega}f_{ele}\ dV=\int_{\Omega}\frac{1}{2}\nabla\phi\cdot\bP\ dV.
$$
We adopt the same idea used before to calculate the $Fr\acute{e}chet$ derivative of the electrostatic energy in this case. For the $Fr\acute{e}chet$ derivative of electrostatic energy at $\bP$ along the direction $\vec{Q}$, we can use the same expression as given in equation \cref{eq:frechet1}. But in this case, $\phi_1$ is replaced by $\phi$, and $\phi(\vec{Q})$ means the solution of the Poisson equation subject to the Neumann boundary condition for a given $\vec{Q}$, i.e., $\Delta\phi(\vec{Q})= \frac{1}{\epsilon}div(\vec{Q})$ in the domain, and $\epsilon\nabla\phi(\vec{Q})\cdot\vec{n}=\vec{Q}\cdot\vec{n}$ on the top and bottom boundary surface (periodicity in the film plane) where $\vec{n}$ is the unit vector normal to the boundary surface. 

For the integral of $\nabla \phi(\vec{Q})\cdot\bP$, we note that the equation \cref{eq:1} still holds. By using the divergence theorem, Green's identity and the periodicity 
of $\bP$, $\phi(\bP)$, $\vec{Q}$ and $\phi(\vec{Q})$ in the X-Y plane as well as the Neumann BC for the two pairs on the top and bottom surface boundary, respectively, we get
\begin{equation}
\label{eq:lqp}
\begin{aligned}
&\int_{\Omega}\nabla \phi(\vec{Q})\cdot\bP\ dV =\int_{\partial\Omega}\phi(\vec{Q})\bP\cdot dS - \int_{\Omega} \phi(\vec{Q})
div \bP dV \\
&=\int_{\partial\Omega}\phi(\vec{Q})\bP\cdot dS+\epsilon(\int_{\Omega}\nabla\phi(\vec{Q})\cdot \nabla\phi(\bP)dV-\int_{\partial\Omega}\phi(\vec{Q})\nabla\phi(\bP)\cdot dS) \\
&=\int_{\partial\Omega}[\phi(\vec{Q})(\bP-\epsilon\nabla\phi(\bP))]\cdot dS+\epsilon(\int_{\partial\Omega}\phi(\bP)\nabla\phi(\vec{Q})\cdot dS-\int_{\Omega}\phi(\bP)\Delta\phi(\vec{Q})\ dV)\\
&=\epsilon \int_{\partial\Omega}\phi(\bP)\nabla\phi(\vec{Q})\cdot dS-\int_{\Omega}\phi(\bP)div\vec{Q}\ dV \\
&=\int_{\partial \Omega}[\phi(\bP)(\epsilon\nabla\phi(\vec{Q})-\vec{Q})]\cdot dS+\int_{\Omega}\nabla \phi(\bP)\cdot\vec{Q}\ dV\\
&=\int_{\Omega}\nabla \phi(\bP)\cdot\vec{Q}\ dV
\end{aligned}
\end{equation}
where $dS$ denotes the normal surface area element, and from the derivation above, combing the $Fr\acute{e}chet$ derivative in equation \cref{eq:frechet1}, we can easily get that the variation of the energy form \cref{eq:electric3}, under the open circuit BC is 
\begin{equation}
\frac{\partial F_{ele} }{\delta \bP}=\nabla\phi(\bP).
\end{equation}
It should be noted that under the Neumann BC, the solutions of the Poisson equation differ from each other by a constant. This is unimportant in general, as the goal is to obtain the electric field $\nabla\phi$ that is unique according to the Poisson equation in this case when given a polarization vector $\bP$.  

\section*{Acknowledgments}
We would like to thank Prof. Long-Qing Chen, Dr. Yu-Lan Li, and Dr. Bo Wang for fruitful discussions. Ruotai~Li also acknowledges the funding support from the China Scholarship Council No.201806010041.

\bibliographystyle{siamplain}
\bibliography{references}

\begin{thebibliography}{10}

\bibitem{Britson_Actamateria_2016_phase}
{\sc J.~Britson, P.~Gao, X.~Q. Pan, and L.~Q. Chen}, {\em Phase field
  simulation of charged interface formation during ferroelectric switching},
  Acta Materialia, 112 (2016), pp.~285--294,
  \url{https://doi.org/10.1016/j.actamat.2016.04.026}.

\bibitem{Chen_JACS_2008_review}
{\sc L.-Q. Chen}, {\em Phase-field method of phase transitions/domain
  structures in ferroelectric thin films: A review}, Journal of the American
  Ceramic Society, 91 (2008), pp.~1835--1844,
  \url{http://dx.doi.org/10.1111/j.1551-2916.2008.02413.x}.

\bibitem{Chen_ComPhyComm_1998_apply}
{\sc L.~Q. Chen and J.~Shen}, {\em Application of semi-implicit
  fourier-spectral method to phase field equations}, Comput. Phys. Commun, 108
  (1998), pp.~147--158, \url{https://doi.org/10.1016/S0010-4655(97)00115-X}.

\bibitem{Choi_Science_2004_enhance}
{\sc K.~J. Choi, M.~Biegalski, Y.~L. Li, A.~Sharan, J.~Schubert, R.~Uecker,
  P.~Reiche, Y.~B. Chen, X.~Q. Pan, V.~Gopalan, L.~Q. Chen, D.~G. Schlom, and
  C.~B. Eom}, {\em Enhancement of ferroelectricity in strained batio3 thin
  films}, Science, 306 (2004), p.~1005,
  \url{https://doi.org/10.1126/science.1103218}.

\bibitem{Choudhury_APL_2008_effect}
{\sc S.~Choudhury, J.~X. Zhang, Y.~L. Li, L.~Q. Chen, Q.~X. Jia, and S.~V.
  Kalinin}, {\em Effect of ferroelastic twin walls on local polarization
  switching: Phase-field modeling}, Applied Physics Letters, 93 (2008),
  p.~162901, \url{http://dx.doi.org/10.1063/1.2993330}.

\bibitem{Cross_Basel_1993}
{\sc L.~Cross}, {\em Ferroelectric ceramics: Tailoring properties for specific
  applications}, in Ferroelectric Ceramics, N.Setter and E.~Colla, eds.,
  Basel(Swizerland):Birkhauser Verlag, 1993, pp.~1--85,
  \url{http://cds.cern.ch/record/113309}.

\bibitem{Dragan_RepProgPhy_1998}
{\sc D.~Damjanovic}, {\em Ferroelectric, dielectric and piezoelectric
  properties of ferroelectric thin films and ceramics}, Reports on Progress in
  Physics, 61 (1998), p.~1267,
  \url{https://iopscience.iop.org/article/10.1088/0034-4885/61/9/002}.

\bibitem{Du_ComMatSci_2009_a}
{\sc Q.~Du and L.~Zhang}, {\em A constrained string method and its numerical
  analysis}, Commun. Math. Sci., 7 (2009), pp.~1039--1051,
  \url{https://projecteuclid.org/euclid.cms/1264434143}.

\bibitem{E_PhyRB_2002_string}
{\sc W.~E, W.~Ren, and E.~Vanden-Eijnden}, {\em String method for the study of
  rare events}, Physical Review B, 66 (2002),
  \url{http://dx.doi.org/10.1103/PhysRevB.66.052301}.

\bibitem{E_JChemP_2007_simplify}
{\sc W.~E, W.~Ren, and E.~Vanden-Eijnden}, {\em Simplified and improved string
  method for computing the minimum energy paths in barrier-crossing events}, J.
  Chem. Phys, 126 (2007), p.~164103, \url{http://dx.doi.org/10.1063/1.2720838}.

\bibitem{Griffiths_Prentice_1999}
{\sc D.~J. Griffiths}, {\em Introduction to Electrodynamics}, Prentice Hall,
  3~ed., 1999, ch.~4, pp.~160--202, \url{https://olin.tind.io/record/123688}.

\bibitem{Haeni_Nature_2004_room}
{\sc J.~H. Haeni, P.~Irvin, W.~Chang, R.~Uecker, P.~Reiche, Y.~L. Li,
  S.~Choudhury, W.~Tian, M.~E. Hawley, B.~Craigo, A.~K. Tagantsev, X.~Q. Pan,
  S.~K. Streiffer, L.~Q. Chen, S.~W. Kirchoefer, J.~Levy, and D.~G. Schlom},
  {\em Room-temperature ferroelectricity in strained srtio3}, Nature, 430
  (2004), pp.~758--761, \url{https://doi.org/10.1038/nature02773}.

\bibitem{Haertling_JACS_1999_ferro}
{\sc G.~H. Haertling}, {\em Ferroelectric ceramics: History and technology},
  Journal of the American Ceramic Society, 82 (1999), p.~797,
  \url{https://doi.org/10.1111/j.1151-2916.1999.tb01840.x}.

\bibitem{Henkelman_JChemP_2000_a}
{\sc G.~Henkelman and H.~Jonsson}, {\em A climbing image nudged elastic band
  method for finding saddle points and minimum energy paths}, J. Chem. Phys,
  113 (2000), \url{https://doi.org/10.1063/1.1329672}.

\bibitem{Khachaturyan_Wiely_1983}
{\sc A.~G. Khachaturyan}, {\em Theory of Structural Transformations in Solids},
  Wiely,New York, 1983, \url{https://www.osti.gov/biblio/5821133}.

\bibitem{Levitt_SIAMnumer_2017}
{\sc A.~LEVITT and C.~ORTNER}, {\em Convergence and cycling in walker-type
  saddle search algorithms}, SIAM J. Numer. Anal., 55 (2017), pp.~2204--2227,
  \url{https://doi.org/10.1137/16M1087199}.

\bibitem{Li_Actamateria_2002_subs}
{\sc Y.~Li, S.~Hu, Z.~Liu, and L.~Chen}, {\em Effect of substrate constraint on
  the stability and evolution of ferroelectric domain structures in thin
  films}, Acta Materialia, 50 (2002),
  \url{https://doi.org/10.1016/S1359-6454(01)00360-3}.

\bibitem{Li_JAP_2004_ferro}
{\sc Y.~L. Li, L.~Q. Chen, G.~Asayama, D.~G. Schlom, M.~A. Zurbuchen, and S.~K.
  Streiffer}, {\em Ferroelectric domain structures in srbi2nb2o9 epitaxial thin
  films: Electron microscopy and phase-field simulations}, Journal of Applied
  Physics, 95 (2004), pp.~6332--6340,
  \url{http://dx.doi.org/10.1063/1.1707211}.

\bibitem{MELines_oxford_1977}
{\sc M.~Lines and A.~Glass}, {\em Principles and applications of ferroelectrics
  and related materials}, Oxford University Press, 1977,
  \url{http://cds.cern.ch/record/367846}.

\bibitem{Pertsev_PRL_2000_polar}
{\sc N.~A. Pertsev and V.~G. Koukhar}, {\em Polarization instability in
  polydomain ferroelectric epitaxial thin films and the formation of
  heterophase structures}, Physical Review Letters, 84 (2000), p.~3722,
  \url{https://doi.org/10.1103/PhysRevLett.84.3722}.

\bibitem{Sepliarsky_JAP_2002_ferro}
{\sc M.~Sepliarsky, S.~R. Phillpot, M.~G. Stachiotti, and R.~L. Migoni}, {\em
  Ferroelectric phase transitions and dynamical behavior in knbo3/ktao3 knbo 3
  / ktao 3 superlattices by molecular-dynamics simulation}, Journal of Applied
  Physics, 91 (2002), p.~3165, \url{https://doi.org/10.1063/1.1435826}.

\bibitem{Speck_JAP_1994_domain}
{\sc J.~S. Speck and W.~Pompe}, {\em Domain configurations due to multiple
  misfit relaxation mechanisms in epitaxial ferroelectric thin films. i.
  theory}, Journal of Applied Physics, 76 (1994), p.~466,
  \url{https://doi.org/10.1063/1.357097}.

\bibitem{Speck_JAP_1994_domain2}
{\sc J.~S. Speck, A.~Seifert, W.~Pompe, and R.~Ramesh}, {\em Domain
  configurations due to multiple misfit relaxation mechanisms in epitaxial
  ferroelectric thin films. ii. experimental verification and implications},
  Journal of Applied Physics, 76 (1994), p.~477,
  \url{https://doi.org/10.1063/1.357098}.

\bibitem{Tenne_PhyRB_2004_abs}
{\sc D.~A. Tenne, X.~X. Xi, Y.~L. Li, L.~Q. Chen, A.~Soukiassian, M.~H. Zhu,
  A.~R. James, J.~Lettieri, D.~G. Schlom, W.~Tian, and X.~Q. Pan}, {\em Absence
  of low-temperature phase transitions in epitaxial batio3 thin films},
  Physical Review B, 69 (2004), p.~174101,
  \url{https://doi.org/10.1103/PhysRevB.69.174101}.

\bibitem{Wang_Actamateria_2007_phase}
{\sc J.~Wang and T.~Y. Zhang}, {\em Phase field simulations of polarization
  switching-induced toughening in ferroelectric ceramics}, Acta Materialia, 55
  (2007), pp.~2465--2477,
  \url{http://dx.doi.org/10.1016/j.actamat.2006.11.041}.

\bibitem{Zhang_SIAMscicomp_2009_diffuse}
{\sc J.~Zhang and Q.~Du}, {\em Numerical studies of discrete approximations to
  the allen--cahn equation in the sharp interface limit}, SIAM J. Sci. Comput,
  31 (2009), pp.~3042--3063, \url{https://doi.org/10.1137/080738398}.

\bibitem{Zhang_SIAMnumer_2012_shrink}
{\sc J.~Zhang and Q.~Du}, {\em Shrinking dimer dynamics and its applications to
  saddle point search}, SIAM J. Numer. Anal., 50 (2012), pp.~1899--1921,
  \url{https://doi.org/10.1137/110843149}.

\bibitem{Zhang_JAP_2008_computer}
{\sc J.~X. Zhang, Y.~L. Li, S.~Choudhury, L.~Q. Chen, Y.~H. Chu, F.~Zavaliche,
  M.~P. Cruz, R.~Ramesh, and Q.~X. Jia}, {\em Computer simulation of
  ferroelectric domain structures in epitaxial bifeo3 thin films}, Journal of
  Applied Physics, 103 (2008), p.~094111,
  \url{http://dx.doi.org/10.1063/1.2927385}.

\bibitem{Zhang_PRL_2007_morph}
{\sc L.~Zhang, L.-Q. Chen, and Q.~Du}, {\em Morphology of critical nuclei in
  solid-state phase transformations}, Physical Review Letters, 98 (2007),
  \url{http://dx.doi.org/10.1103/PhysRevLett.98.265703}.

\bibitem{Zhang_JCP_2010_diffuse}
{\sc L.~Zhang, L.-Q. Chen, and Q.~Du}, {\em Diffuse-interface approach to
  predicting morphologies of critical nucleus and equilibrium structure for
  cubic to tetragonal transformations}, J. Comput. Phys., 229 (2010),
  pp.~6574--6584, \url{http://dx.doi.org/10.1016/j.jcp.2010.05.013}.

\bibitem{Zhang_SIAMJSCom_2016_opt}
{\sc L.~Zhang, Q.~Du, and Z.~Zheng}, {\em Optimization-based shrinking dimer
  method for finding transition states}, SIAM J. Sci. Comput, 38 (2016),
  pp.~528--544, \url{https://doi.org/10.1137/140972676}.

\bibitem{Zhang_npj_2016_recent}
{\sc L.~Zhang, W.~Ren, A.~Samanta, and Q.~Du}, {\em Recent developments in
  computational modelling of nucleation in phase transformations}, Npj
  Computational Materials, 2 (2016),
  \url{https://doi.org/10.1038/npjcompumats.2016.3}.

\end{thebibliography}
\end{document}